%% file: scalar.tex
\documentclass[12pt]{article}
\usepackage{epsfig}

\input econfmacros.tex

\def\kpipi{$D^+\rightarrow K^-\pi^+\pi^+$\ }
\def\kpp{$K^-\pi^+\pi^+$\ }
\def\ka14{$K^*_0(1430)$\ }
\def\d3pi{$D^+\rightarrow \pi^-\pi^+\pi^+$\ }
\def\ds3pi{$D_s^+\rightarrow \pi^-\pi^+\pi^+$\ }
\def\Title#1{\begin{center} {\Large {\bf #1} } \end{center}}

\begin{document}

\Title{Light  Scalar Mesons $\sigma(500)$, $f_0(980)$ and $\kappa$ in Charm Meson 
Decays \footnote{To appear in proceedings of VIII International Workshop on Hadron Physics 2002\\
 Rio Grande do Sul, Brazil.  World Scientific edition.}}

\bigskip\bigskip

\begin{raggedright}  

{{\it Ignacio Bediaga} \\
 Representing the Fermilab E791 Collaboration \\
Centro Brasileiro de Pesquisas F\'\i sicas, \\
Rua Xavier Sigaud 150, 22290, Rio de Janeiro, Brazil \\
bediaga@cbpf.br}

\bigskip\bigskip

\end{raggedright}

\begin{abstract}
We present recent results on scalar light mesons based on Dalitz plot 
analyses of charm decays from Fermilab experiment E791. Scalar mesons are found
to have large contributions to the decays studied, $D^+\to K^-\pi^+\pi^+$ and 
$D^+, D_s^+\to\pi^-\pi^+\pi^+$. From the first decay,  we find good 
evidence for the existence of the light and broad $\kappa$ meson and we measure 
its mass and width. We find strong evidence for the $\sigma(500)$ meson 
from $D^+\to\pi^-\pi^+\pi^+$ decay and measure  its mass and width. We also 
present the results obtained for the $f_0(980)$ parameters through the 
$D^+_s \to\pi^-\pi^+\pi^+$ decay.  These results demonstrate the importance  
of charm decays as a new environment for the study of light meson physics.

\end{abstract}

\section{Introduction}

The decays of charm mesons are currently a new source of information for 
the study of light meson spectroscopy, with the advantages of having well 
defined initial state (the $D$ meson, a $0^-$ state with defined mass). 
This new information is complementary to that from scattering experiments 
and can be particularly relevant to the understanding of the scalar sector, 
due to the well-known difficulties of scattering experiments to observe 
light and broad scalar resonances.

Here we present an overview of the  results we obtained  analysing  
the decays \ds3pi \cite{ds3pi}, \d3pi \cite{d3pi} and  \kpipi \cite{kpipi}, 
using data collected in 1991/92 by  Fermilab experiment E791   from 
500 GeV/c $\pi^-$--nucleon interactions. For details see \cite{ref791}. 

For the \d3pi decays, we find that a model with only known $\pi\pi$ resonances 
plus a non-resonant (NR) channel is not able to describe the data adequately. 
Thus we include 
a new amplitude in our fit function and find  strong evidence for the
presence of a light and broad scalar resonance, the $\sigma(500)$. The channel 
involving  this scalar meson is responsible for half of the decay rate. We 
measure 
the mass and the width of this scalar meson to be $ 478^{+24}_{-23} \pm 17  
$ MeV/$c^2$ and $ 324^{+42}_{-40}\pm 21$ MeV/$c^2$,  respectively. Other 
experiments have presented controversial evidence for this 
low-mass $ \pi \pi $ resonance in partial wave analyses \cite{wa102,cleo,gams},
with ambiguous interpretations of  the characteristics of such 
particles\cite {pdg,torn1}.

We found a similar situation for the \kpipi analysis. When we include all 
known $K\pi$ resonant channels plus a NR contribution, we find that 
this model is not able to describe the data.  By including an extra
scalar resonant state, with unconstrained mass and width, we obtain a fit which is substantially
superior to that without this state. The values for its mass and width are found to be
$797\pm 19\pm 42$ MeV/c$^2$ and $410\pm 43\pm 85$ MeV/c$^2$ respectively.
We refer to this state as the $\kappa$. We also obtain new measurements 
for the mass and the width of the \ka14 resonance.

From our analysis of \ds3pi decays we found that the  isoscalar intermediate states 
are dominant in  this decay as well.  The largest  
contribution to this final state comes from the decay involving  
the scalar meson $f_0(980)$ whose nature  is a long-standing puzzle. 
It has been described as a $q \bar q$ state , a $K \bar K$ molecule,
 a glueball, and  a multiquark state \cite{pdg,torn1}. We obtain new measurements 
for the $f_0(980)$ and $f_0(1370)$ masses and widths.

To obtain these results, we had to introduce a new approach in Dalitz plot 
analysis in order to extract the mass and width of the scalar resonances 
by using them as floating parameters in the fit.  
We begin this paper presenting the general  method, applied in the \kpipi 
Dalitz-plot analysis, then we discuss the \ds3pi and \d3pi studies using the 
same procedure.

\section{The \kpipi Dalitz-plot Analysis}

\begin{figure}[t]
\centerline{
\begin{minipage}{2.5in}
\epsfxsize=13pc
\epsfbox{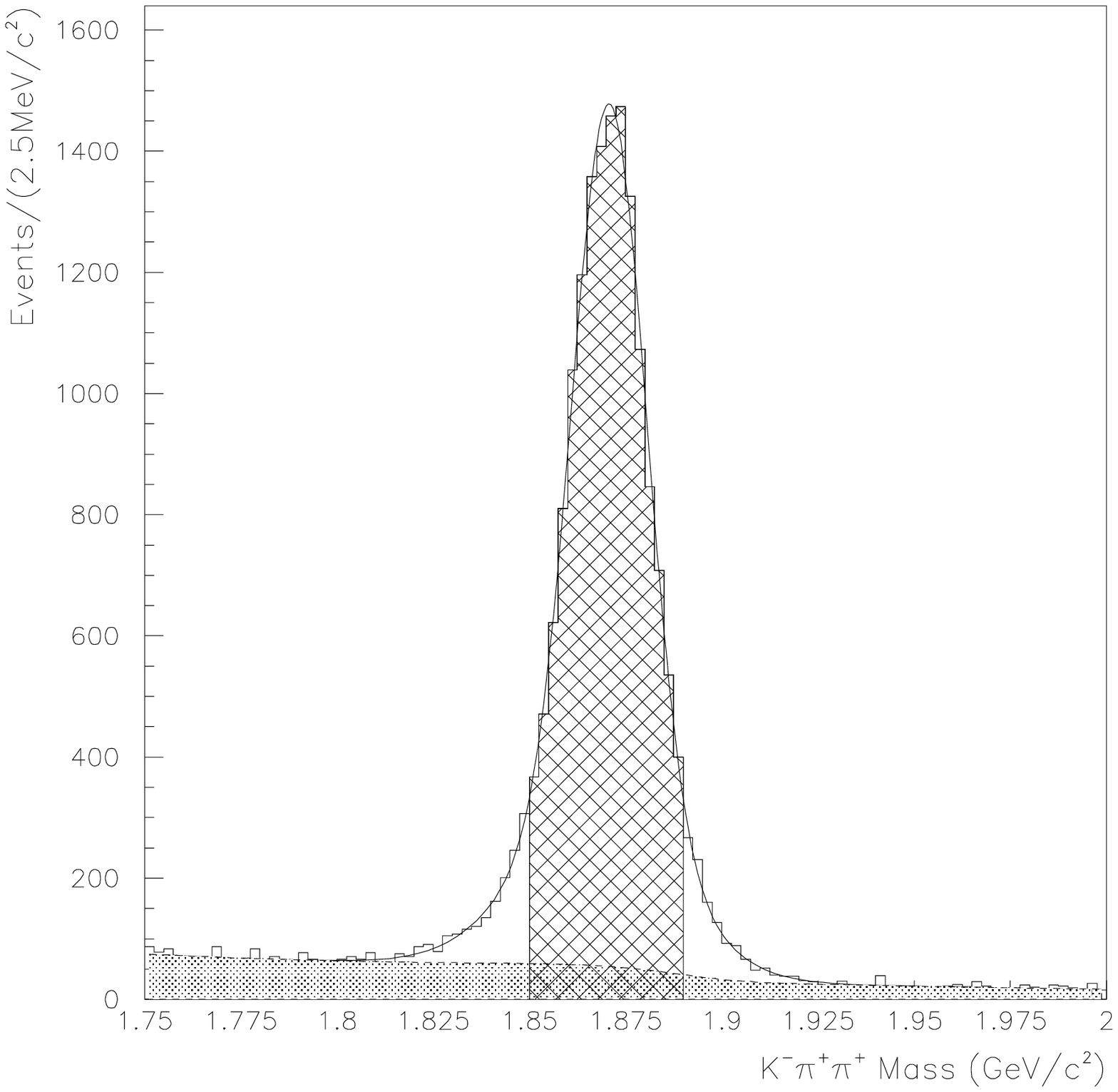}
\end{minipage}
\begin{minipage}{2.5in}
\epsfxsize=13pc
\epsfbox{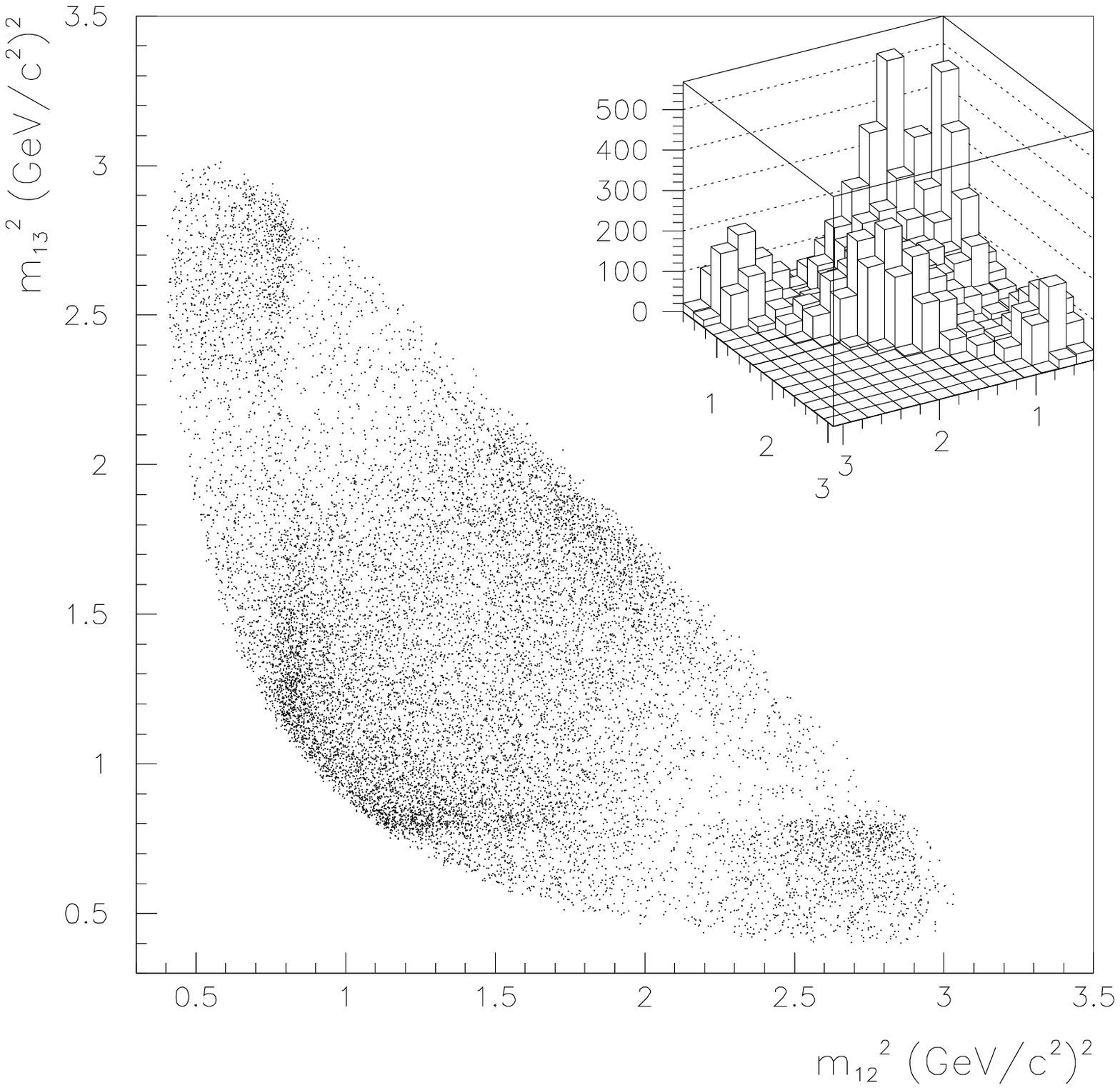}
\end{minipage}}
\caption{(a) The \kpp invariant mass spectrum. The filled area is 
background; (b) Dalitz plot corresponding to the events in the cross-hatched area of (a).
\label{kpipi} }
\end{figure}

From the original $2\times 10^10$ events collected by E791, and after reconstruction and
selection criteria, we obtained the \kpipi sample shown in Figure~\ref{kpipi}(a). 
The filled area represents the level of background; besides the combinatorial, 
the other main source 
of background comes from the reflection of the decay $D_s^+\to K^-K^+\pi^+$ 
(through $\bar K^*K^+$ 
and $\phi\pi^+$). 
The cross-hatched region contains the events selected for the Dalitz-plot ana\-lysis. 
There are 15090 events in this sample, of which 6\% are background. 

Figure~\ref{kpipi}(b) shows the Dalitz-plot for these events. The two axes are the squared 
invariant-mass combinations for $K\pi$, and the plot
is symmetrized with respect to the two identical pions.
The plot presents a rich structure, where we can observe the clear
bands from $\bar K^*(890)\pi^+$, and an accumulation of events at the upper edge of
the diagonal, due to heavier resonances.
To study the resonant substructure, we perform an unbinned maximun-likelihood fit to the 
data, with probability distribution functions (PDF's) for both signal and background sources.
In particular, for each candidate event, the signal PDF is written as the square of the 
total physical amplitude ${\cal A}$  and it is weighted
for the acceptance across the Dalitz plot (obtained by Monte Carlo (MC)) and by 
the level of signal to background for each event, as given by the line shape of 
Figure~\ref{kpipi}(a). The background PDF's (levels and shapes) are fixed for the Dalitz-plot fit, 
according to MC and data studies.

We begin describing our first approach to fit the data, which represents the conventional
Dalitz-plot analysis including the known $K\pi$ resonant amplitudes (${\cal A}_n,~n\ge 1$), plus 
a constant non-resonant contribution. The signal amplitude, a Breit-Wigner
parametrization times a Blatt-Weisskopf damping factor, is described 
in reference \cite{kpipi}.

Using this model with well-known resonances (Model A), we find contributions from the following channels: the non-resonant,
responsible for more than 90\% of the decay rate, followed by $\bar K^*_0(1430)\pi^+$, 
$\bar K^*(892)\pi^+$, $\bar K^*(1680)\pi^+$ and $\bar K^*_2(1430)\pi^+$. The decay fractions 
and relative phases are shown in Table~\ref{tablekappa}. 
These values are in accordance with previous results from E691 \cite{e691-kpipi} and E687
\cite{e687-kpipi}. We thus confirm a high
non-resonant contribution according to this model, which is totally unusual in $D$ decays.
Besides, there is an important destructive interference pattern, since all fractions add up
to 140 \%.

\begin{figure}[t]
\centerline{
\begin{minipage}{2.5in}
\epsfxsize=15pc
\epsfbox{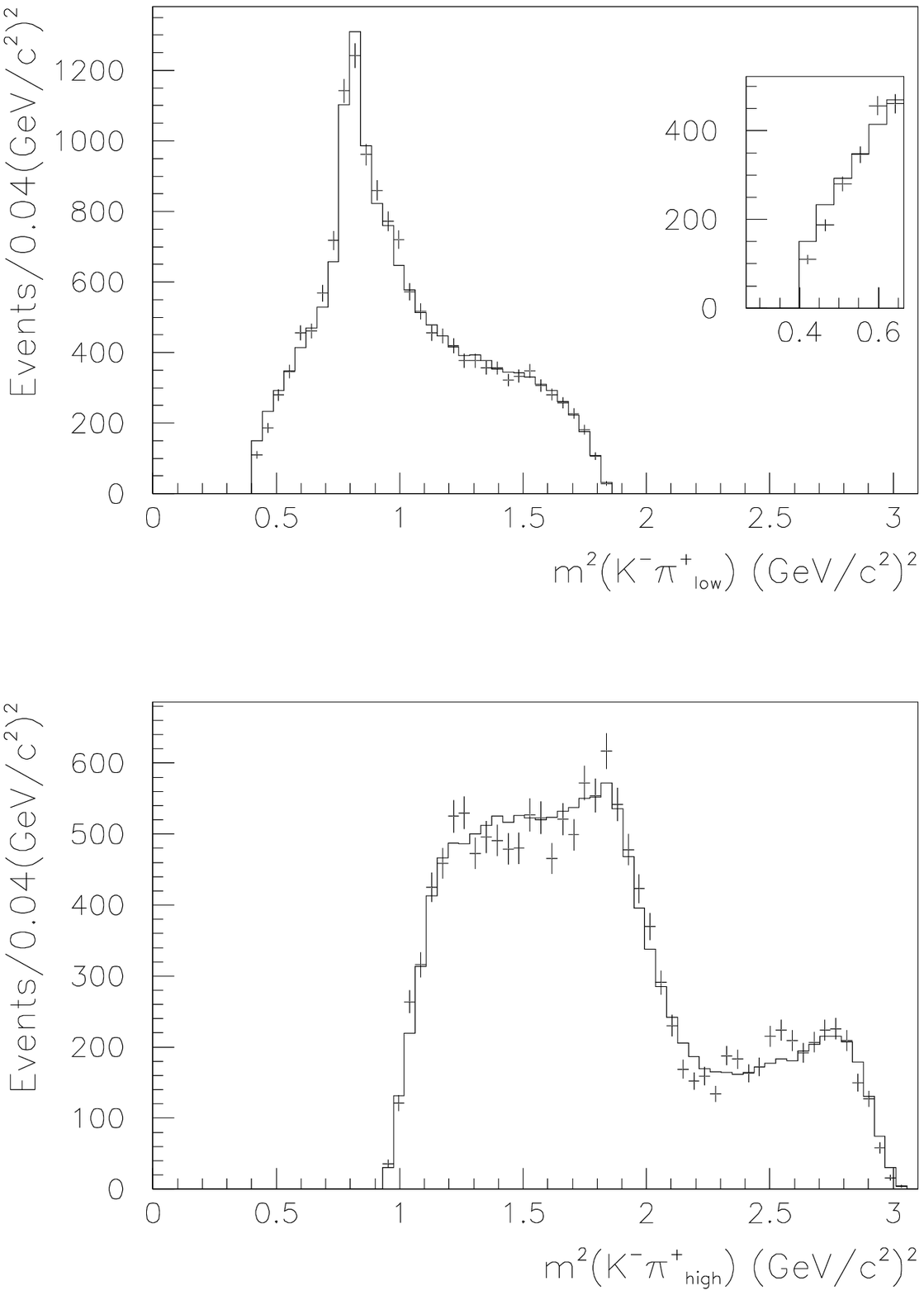}
\end{minipage}
\begin{minipage}{2.5in}
\epsfxsize=15pc
\epsfbox{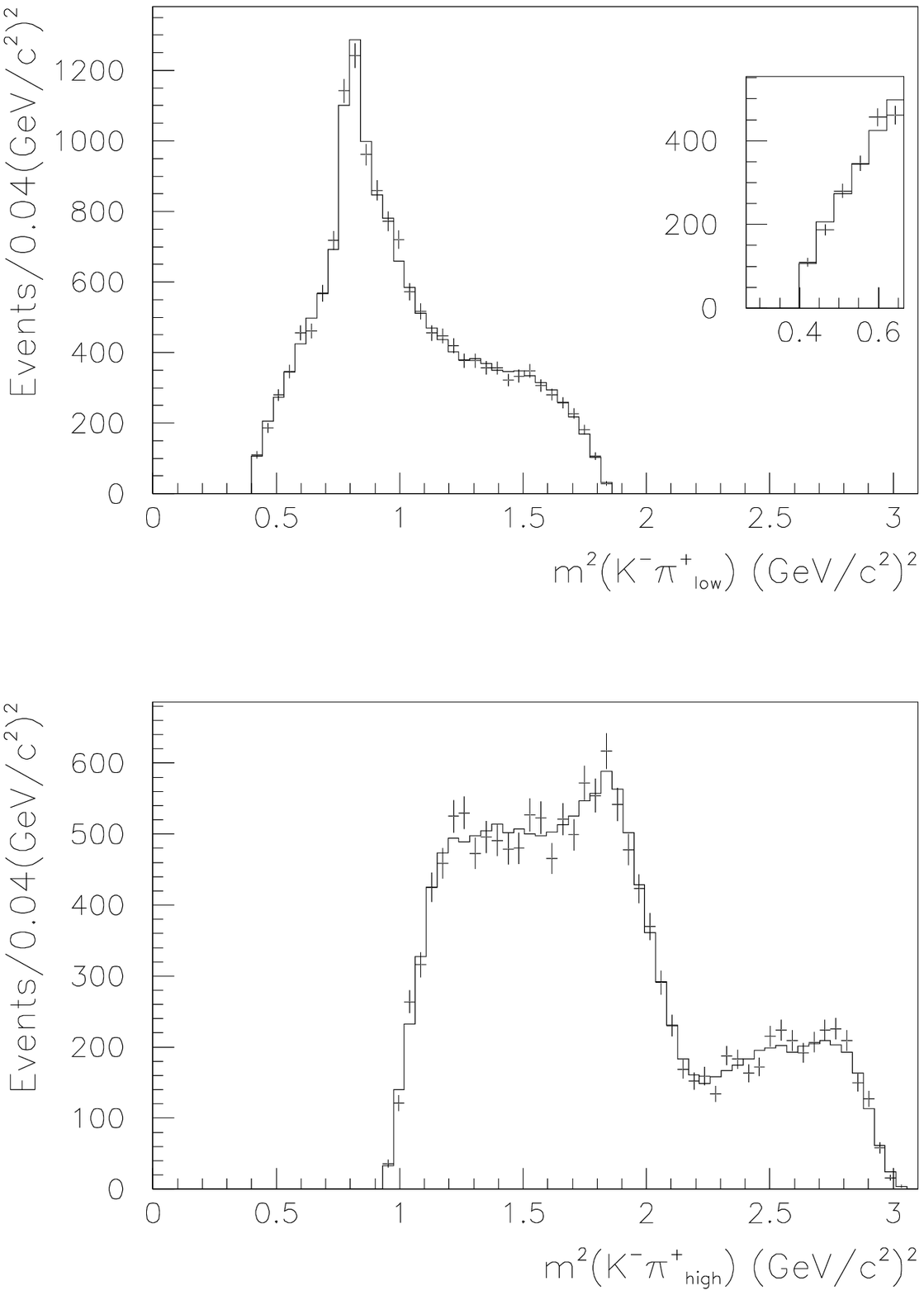}
\end{minipage}}
\caption{$m^2(K\pi_{\rm low})$ and $m^2(K\pi_{\rm high})$ projections for data
(error bars) and fast MC (solid line): (a) fit to Model A, without $\kappa$, and (b) fit to Model B,
with $\kappa$. }
\label{proj_kpipi} 
\end{figure}

To evaluate the fit quality, we compute a $\chi^2$ from binned, 
two-dimensional distributions of data and decay model events. 
The $\chi^2$ comes from the differences in the binned numbers of events between
the data and the model (from a fast MC simulation).
We obtain $\chi^2/\nu=2.7$ ($\nu$ being the number of degrees of freedom), with a corresponding 
confidence level (CL) of $10^{-11}$. In Figure~\ref{proj_kpipi}(a) we show the $K\pi$ low and
high squared-mass projections for data (error bars) and model (solid line). 
The discrepancies are evident in the very low-mass region for $m^2(K\pi)_{\it low}$ 
and near 2.5 (GeV/c$^2$)$^2$ for  $m^2(K\pi)_{\it high}$. These regions of disagreement 
are the same observed previously by E687 \cite{e687-kpipi}. We thus conclude 
that a model with the known $K\pi$ resonances, plus a non-resonant amplitude, 
is not able to describe the \kpipi Dalitz plot satisfactorily.

A similar pattern -- bad fit quality with large NR fraction -- is found in the analysis of the
decay \d3pi when allowing only the established $\pi\pi$ resonances \cite{d3pi} .
There we find that the inclusion of an extra scalar resonance improves the fit substantially, giving
strong evidence for the $\sigma(500)$. See the section on \d3pi below.
Thus, we are led to try an extra scalar resonance in our fit model here.

This second fit model, Model B, is constructed by the 
inclusion of an extra scalar state, with unconstrained mass
and width. For consistency, the mass and width of the other scalar state, the $K^*_0(1430)$, are 
also free parameters of the fit. We adopt a better description for these scalar states by 
introducing gaussian-type form-factors \cite{torn2} to take into account the finite size of the
decaying mesons. Two extra floating parameters are the meson radii 
$r_D$ and $r_R$ introduced for these meson sizes.

Using this model, we obtain the values of  $797\pm 19\pm 42$ MeV/c$^2$ for the mass and 
$410\pm 43\pm 85$ MeV/c$^2$ for the width of the new scalar state (first error statistical, 
second error systematic), referred to here as the $\kappa$.
The values of mass and width obtained for the $K^*_0(1430)$ are respectively 
$1459\pm 7\pm 6$ MeV/c$^2$ and $175\pm 12\pm 12$ MeV/c$^2$, appearing heavier and
narrower than presented by the PDG \cite{pdg}. The decay fractions and 
relative phases for 
Model B, with statitical errors, are given in Table~\ref{tablekappa}. 
Compared to the results of Model A (without $\kappa$), 
the non-resonant mode drops from over 90\% to 13 $\pm$ 6\%. The $\kappa\pi^+$ state 
is now the dominant channel with decay fraction about 50\%. 
The meson radii $r_D$ and $r_R$ are found to be respectively $5.0\pm 0.5$ GeV$^{-1}$ and 
$1.6\pm 1.3$ GeV$^{-1}$, in  agreement with values used by other groups 
 \cite{argus,cleo2}.

\begin{table}\centering
\begin{tabular}{|c|c c|c c|} \hline 
 Decay & \multicolumn{2}{|c|}{Model A: No $\kappa$} 
 & \multicolumn{2}{|c|}{Model B: With $\kappa$ }  \\
 Mode  & Fraction (\%) & Phase & Fraction (\%) & Phase  \\  \hline
NR           & $90.9\pm 2.6$ & $0^\circ$ (fixed) & $13.0\pm 5.8\pm 2.6$ & $(349\pm 14\pm 8)^\circ$  \\ 
$\kappa\pi^+$  & -- & -- & $47.8\pm 12.1\pm 3.7$ & $(187\pm 8\pm 17)^\circ$ \\ 
$\bar K^*(892)\pi^+$  & $13.8\pm 0.5$ & $(54\pm 2)^\circ$ & $12.3\pm 1.0\pm 0.9$ & $0^\circ$ (fixed)  \\ 
$\bar K^*_0(1430)\pi^+$ & $30.6\pm 1.6$ & $(54\pm 2)^\circ$ & $12.5\pm 1.4\pm 0.4$ & $(48\pm 7\pm 10)^\circ$ \\ 
$\bar K^*_2(1430)\pi^+$   & $0.4\pm 0.1$ & $(33\pm 8)^\circ$ & $0.5\pm 0.1\pm 0.2$ & $(306\pm 8\pm 6)^\circ$\\ 
$\bar K^*(1680)\pi^+$    & $3.2\pm 0.3$ & $(66\pm 3)^\circ$ & $2.5\pm 0.7\pm 0.2$ & $(28\pm 13\pm 15)^\circ$\\ \hline 
\end{tabular}
\caption{Results without $\kappa$ (Model A) and with $\kappa$ (Model B). 
\label{tablekappa}}
\end{table}
Moreover, the fit quality of Model B is substantially superior to that of Model A. 
The $\chi^2/\nu$ is now 0.73 with a CL of 95\%. The very good agreement between
the model and the data can be seen in the projections of Figure~\ref{proj_kpipi}(b).
A number of studies were done to check these results. They are described 
in detail in reference \cite{kpipi}.

\section{The \ds3pi Results}

\begin{figure}[t]
\epsfxsize=20pc
\centerline{
\epsfbox{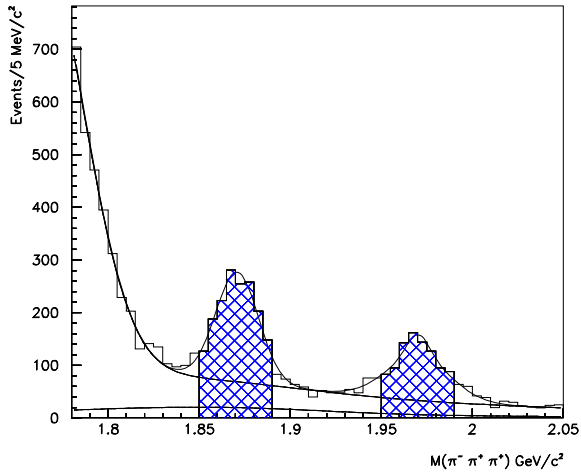}} 
\caption{The $\pi^-\pi^+\pi^+$ invariant mass spectrum. The  dashed line 
represent the total background. Events used for the Dalitz analyses
are in the hatched areas.}
\label{m3pi} 
\end{figure}

\begin{figure}[t]
\centerline{
\begin{minipage}{2.5in}
\epsfxsize=12pc
\epsfbox{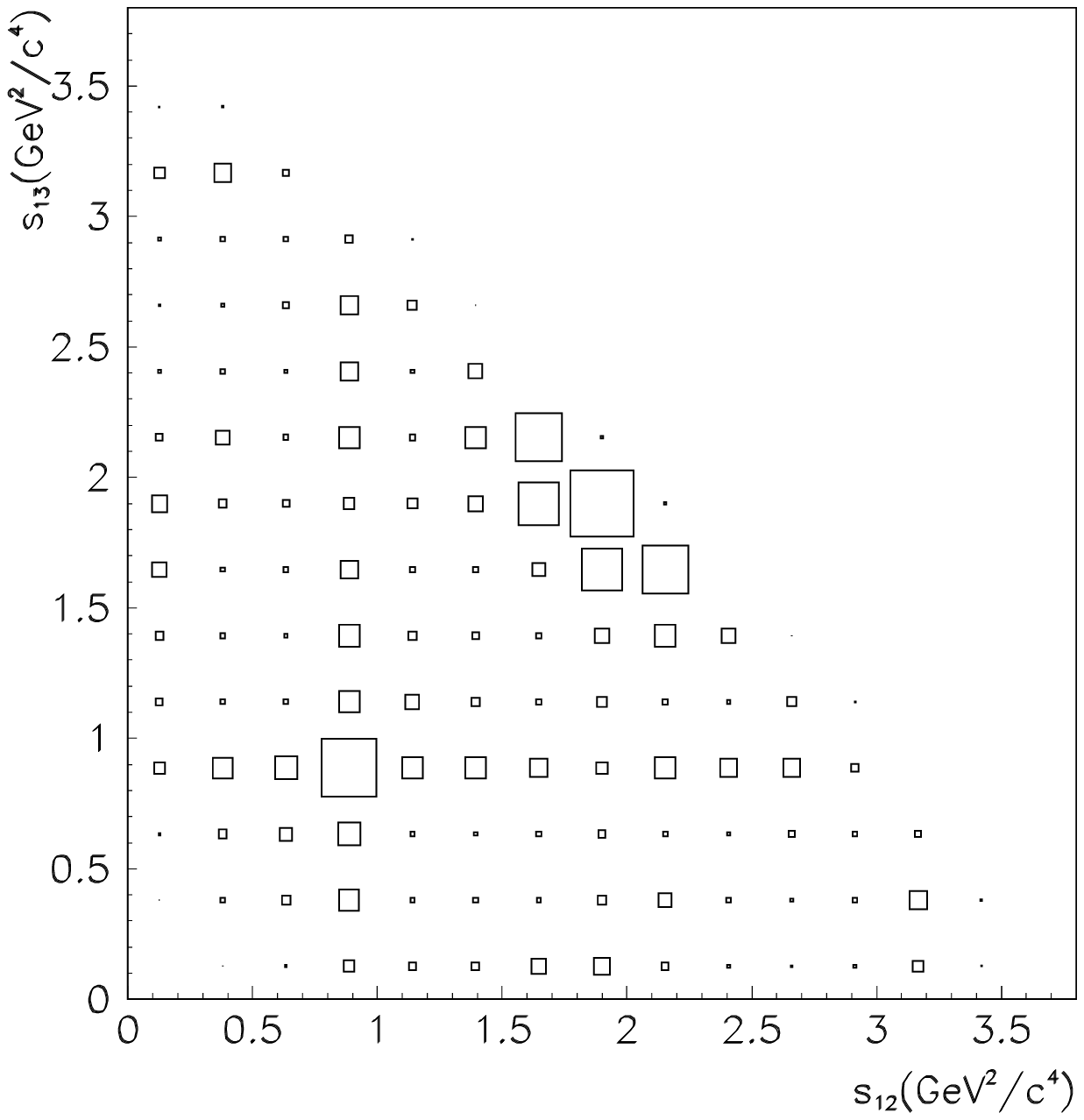}
\end{minipage}
\begin{minipage}{2.5in}
\epsfxsize=12pc
\epsfbox{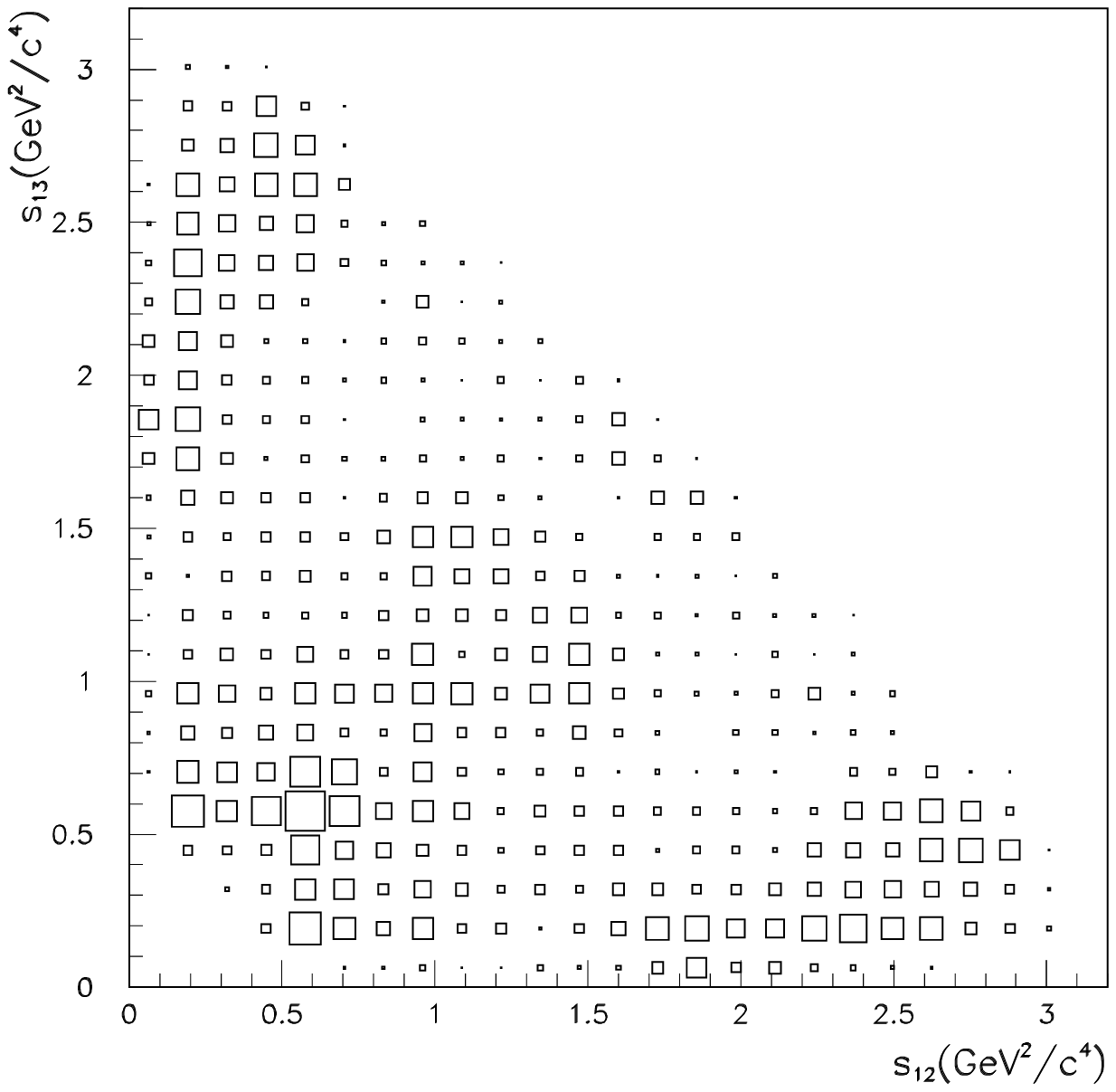}
\end{minipage}}
\caption{(a) The $D_s^+ \to \pi^- \pi^+ \pi^+$ Dalitz plot and
(b) the \d3pi Dalitz plot. Since there are two identical pions, 
the plots are symmetrized.
\label{dalitz3pi} } 
\end{figure}

In Figure~\ref{m3pi} we show the $\pi^-\pi^+\pi^+$ invariant mass distribution 
after reconstruction and selection criteria \cite{ds3pi,d3pi}
for the sample collected by E791. 
Besides combinatorial background, reflections from the decays \kpipi, $D^0\to K^-\pi^+$ 
(plus one extra $\pi^+$ track) and $D_s^+\to \eta'\pi^+,
~\eta'\to\rho^0(770)\gamma$ are all taken into account.
The hatched regions in Figure \ref{m3pi} show the samples used for the
Dalitz-plot analyses. There are 1686 and 937 candidate events for $D^+$  
and $D_s^+$ respectively, with a signal to background ratio of about 2:1. 
The Dalitz plots for these events are shown 
in Figure~\ref{dalitz3pi}, the axes corresponding to the two $\pi^-\pi^+$ invariant-masses 
squared.

For the \ds3pi events in Figure~\ref{dalitz3pi}(a), the signal amplitude includes
all channels with well-known dipion resonances \cite{pdg}: $\rho^0(770) \pi^+$, 
$f_0(980) \pi^+$, $f_2(1270) \pi^+$, $f_0(1370) \pi^+$, $\rho^0(1450) \pi^+$ 
and the non-resonant, assumed constant across the Dalitz plot. 

For the  $f_0(980)\pi^+$ amplitude, instead of a simple Breit-Wigner form,
we use a coupled-channel Breit-Wigner function \cite{wa76},
\begin{equation}
BW_{f_0(980)} = {1 \over {m_{\pi\pi}^2 - m^2_0 + im_0(\Gamma_{\pi}+\Gamma_K)}}\ ,
\end{equation}
\begin{equation}
\Gamma_{\pi} = g_{\pi}\sqrt{m_{\pi\pi}^2/ 4 - m_{\pi}^2},~
\Gamma_K = {g_K \over 2} \left( \sqrt{m_{\pi\pi}^2/ 4 - m_{K^+}^2}+
\sqrt{m_{\pi\pi}^2/ 4 - m_{K^0}^2}\right) .
\end{equation}
The \ds3pi Dalitz plot is fit to obtain not only the 
decay fractions and phases of the possible sub-channels, but also 
the parameters of the $f_0(980)$ state, $g_{\pi}$, $g_K$, and $m_0$, as well 
as the mass and width of the $f_0(1370)$. The other resonance 
masses and widths are taken from the PDG\cite{pdg}.
The resulting fractions and phases are given in Table \ref{tabds3pi}. 
The measured $f_0(980)$ parameters are 
 $m_0 =  977 \pm 3 \pm 2$ MeV/c$^2$, $g_{\pi} =$ 0.09  $\pm$  0.01 $\pm$ 0.01 
and  $g_K =$ 0.02  $\pm$  0.04  $\pm$  0.03. By fitting the Dalitz plot 
using  a simple Breit-Wigner 
function, for the $f_0(980)$ we find $m_0 = 975 \pm 3$ MeV/c$^2$ and 
$\Gamma_0 =  44 \pm 2 \pm 2$ MeV/c$^2$, and the results
for fractions and phases are indistinguishable.

\begin{figure}[htb]
\epsfxsize=25pc
\centerline{
\epsfbox{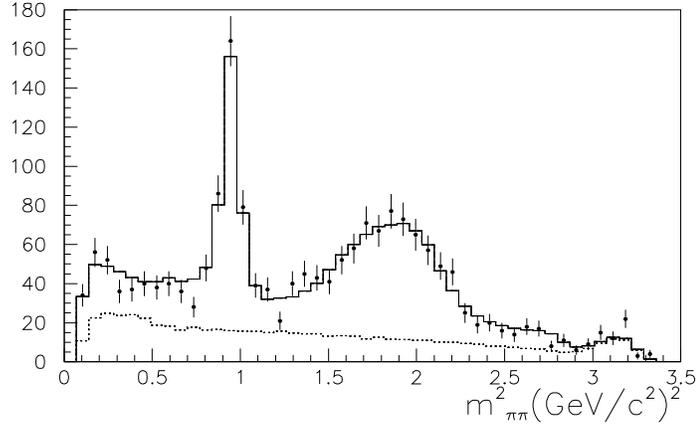}} 
\caption{ $s_{12}$ and $s_{13}$ ($m^2_{\pi\pi}$) projections for 
\ds3pi data (dots) and our best fit
(solid). The  area under the dashed curve corresponds to  background.
\label{proj_ds} } 
\end{figure}

The confidence level of the fit for \ds3pi is 35\%. 
In Figure~\ref{proj_ds} we show
the $\pi^-\pi^+$ mass-squared projections for data (points)
and model (solid lines, from fast-MC).

\begin{table}[htb]\centering
\begin{tabular}{|c|c c|} \hline
Decay Mode          &    Fraction (\%)              &  Phase   \\ \hline 
$f_0(980)\pi^+$     & $56.5\pm 4.3\pm 4.7$ & $0^\circ$ (fixed)   \\ 
NR          & $ 0.5\pm 1.4\pm 1.7$ &~$(181\pm 94\pm 51)^\circ$~\\ 
$\rho^0(770)\pi^+$  & $ 5.8\pm 2.3\pm 3.7$ & $(109\pm 24\pm  5)^\circ$ \\ 
$f_2(1270)\pi^+$    & $19.7\pm 3.3\pm 0.6$ & $(133\pm 13\pm 28)^\circ$ \\ 
$f_0(1370)\pi^+$    & $32.4\pm 7.7\pm 1.9$  & $(198\pm 19\pm 27)^\circ$ \\ 
$\rho^0(1450)\pi^+$ & $ 4.4\pm 2.1\pm 0.2$ & $(162\pm 26\pm 17)^\circ$ \\ \hline
\end{tabular}
\caption{Dalitz fit results for \ds3pi.}
\label{tabds3pi}
\end{table}


\section{The \d3pi Results}
\label{secd3pi}

\begin{figure}[htb]
\centerline{
\begin{minipage}{3.in}
\epsfxsize=18pc
\epsfbox{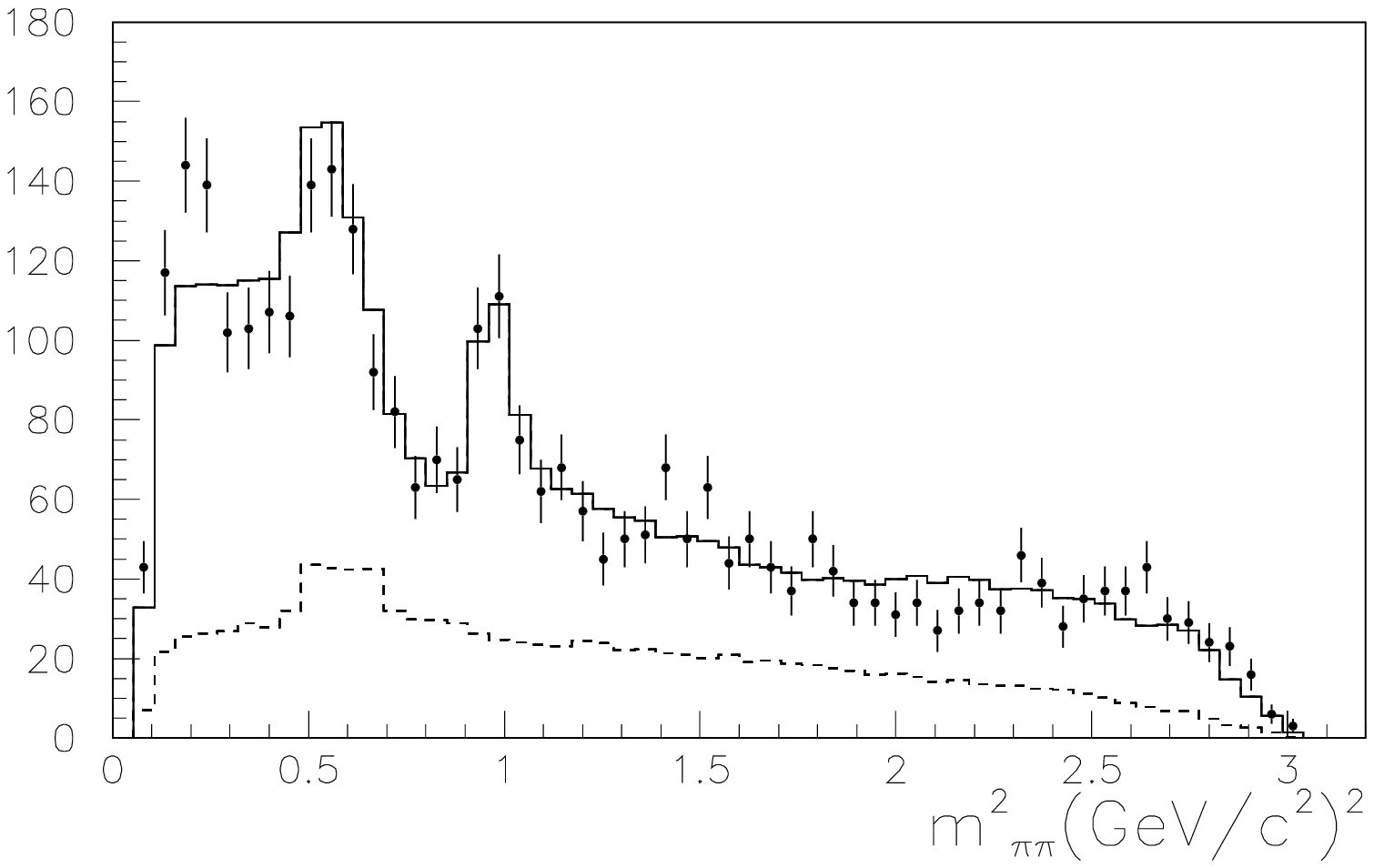}
\end{minipage}
\begin{minipage}{2.5in}
\epsfxsize=18pc
\epsfbox{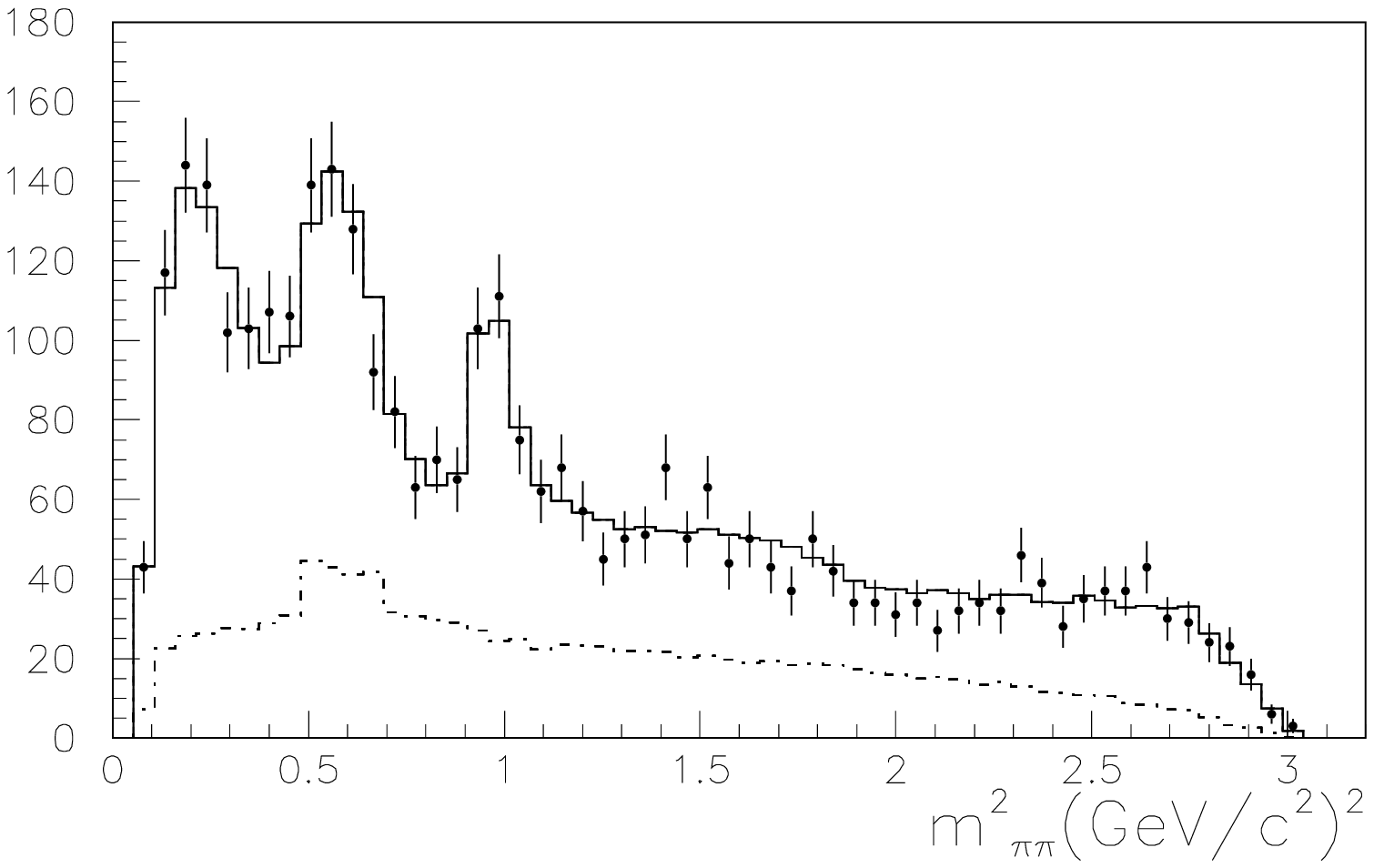}
\end{minipage}}
\caption{$s_{12}$ and $s_{13}$ ($m^2_{\pi\pi}$) projections for \d3pi data (dots) 
and our best fit (solid) for models \rm (a) without and \rm (b) with $\sigma(500)\pi^+$ amplitude. 
The dashed distributions corresponds to the expected background levels.}
\label{proj_dp3pi} 
\end{figure}

In a first approach, we try to fit the \d3pi Dalitz plot of Figure~\ref{dalitz3pi}(b)
with the same amplitudes used for the \ds3pi analysis. Using this model, 
the non-resonant, the $\rho^0(1450)\pi^+$, and the $\rho^0(770)\pi^+$ amplitudes
are found to dominate, as shown in  Table~\ref{tabd3pi}, and in agreement with
previous reported analyses \cite{e691-3pi,e687-3pi}.
However, this model does not describe the data satisfactorily, especially at low 
$\pi^-\pi^+$ mass squared, as can be seen from Fig.~\ref{proj_dp3pi}(a). The $\chi^2/\nu$ 
obtained from the binned Dalitz plot for this model is 1.6, with a CL less 
than $10^{-5}$. 
   
To investigate the possibility that another $\pi^-\pi^+$ resonance contributes to the
\d3pi decay, we add an extra scalar resonance amplitude to the signal PDF, with  
mass and width as floating parameters in the fit.

\begin{table}[htb]\centering
\begin{tabular}{|c|c c|c c|} \hline  
Decay & \multicolumn{2}{|c|}{Fit without $\sigma(500)\pi^+$} 
      & \multicolumn{2}{|c|}{Fit with $\sigma(500)\pi^+$} \\ 
Mode               &    Fraction (\%) &  Phase
                   &    Fraction (\%)  &    Phase  \\ \hline 
$\sigma(500)\pi^+$      &         --               &          -- 
                   & $46.3\pm 9.0\pm 2.1$ & $(206\pm 8\pm 5)^\circ$\\ 
$\rho^0(770)\pi^+$ & $20.8\pm 2.4$  & $0^\circ$ (fixed) 
                   & $33.6\pm 3.2\pm 2.2$ & $0^\circ$ (fixed)     \\ 
NR         & $38.6\pm 9.7$  & $(150\pm 12)^\circ$     
                   & $ 7.8\pm 6.0\pm 2.7$ & $(57\pm 20\pm 6)^\circ$  \\ 
$f_0(980)\pi^+$    & $7.4\pm 1.4$   & $(152\pm 16)^\circ$    
                   & $ 6.2\pm 1.3\pm 0.4$ & $(165\pm 11\pm 3)^\circ$ \\ 
$f_2(1270)\pi^+$   & $6.3\pm 1.9$   & $(103\pm 16)^\circ$   
                   & $19.4\pm 2.5\pm 0.4$ & $(57\pm 8\pm 3)^\circ$  \\ 
$f_0(1370)\pi^+$   & $10.7\pm 3.1$  & $(143\pm 10)^\circ$     
                   & $ 2.3\pm 1.5\pm 0.8$ & $(105\pm 18\pm 1)^\circ$ \\ 
$\rho^0(1450)\pi^+$& $22.6\pm 3.7$  & $ (46\pm 15)^\circ$    
                   & $ 0.7\pm 0.7\pm 0.3$ & $(319\pm 39\pm 11)^\circ$\\ \hline
\end{tabular}
\caption{Dalitz fit results for \d3pi. First errors are statistical, second 
systematics (only for fit with $\sigma(500)\pi^+$ mode).}
\label{tabd3pi}
\end{table}

We find that this model improves our fit substantially. 
The mass and the width of the extra scalar state are found to be 
$ 478^{+24}_{-23} \pm 17  $ MeV/$c^2$ and $ 324^{+42}_{-40}  \pm 21$ MeV/$c^2$, respectively.
Refering to this state as the $\sigma(500)$, we obtain that 
$\sigma(500)\pi^+$ channel produces the largest decay fraction, as shown in 
Table~\ref{tabd3pi}. The non-resonant amplitude, which is dominant in the 
model without $\sigma(500)\pi^+$, drops substantially. 
This model describes the data much better, as can be seen by the $\pi\pi$ 
mass squared projection in Fig.~\ref{proj_dp3pi}(b). The $\chi^2/\nu$ is 
now 0.9, with a corresponding  confidence level of 91\%.

\newpage
\section{Conclusion}

From the data of the Fermilab E791 experiment, we studied the Dalitz plots of the
decays $D^+\to K^-\pi^+\pi^+$, \ds3pi and $D^+\rightarrow \pi^-\pi^+\pi^+$. 
In these three final states, the scalar intermediate resonances were found to give the main
contribution to the decay rates. We obtained strong evidence for the existence of  
$\sigma(500)$ and $\kappa$ scalar mesons, 
measuring their masses and widths. We also obtained new measurements for masses and
widths of the other scalars studied, $f_0(980)$, $f_0(1430)$ and $K^*_0(1430)$.

The results presented here show the potential of $D$ meson decays for the study of 
light meson spectroscopy, in particular in the scalar sector.

\end{document}

%% file: econfmacros.tex
\def\beq{\begin{equation}}
\def\eeq#1{\label{#1}\end{equation}}
\def\eeqn{\end{equation}}

\def\beqa{\begin{eqnarray}}
\def\eeqa#1{\label{#1}\end{eqnarray}}
\def\eeqan{\end{eqnarray}}

\let\bar=\overbar

\def\Dslash{\not{\hbox{\kern-4pt $D$}}}
\def\dslash{\not{\hbox{\kern-2pt $\del$}}}

\def\msb{{\bar{\ssstyle M \kern -1pt S}}}

